# A secure blind watermarking scheme based on DCT domain of the scrambled image

Lei Chen, Shihong Wang*

*School of Sciences, Beijing University of Posts and Telecommunications, Beijing 100876, China*

**Abstract**

This paper investigates a secure blind watermarking scheme. The main idea of the scheme not only protects the watermark information but also the embedding positions. To achieve a higher level of security, we propose a sub key generation mechanism based on the singular value decomposition and hash function, where sub keys depend on both the main key and the feature codes of the original image. The different sub keys ensure that the embedding positions are randomly selected for different original images. Watermark is embedded in the Discrete Cosine Transform (DCT) coefficients of the scrambled original image. Simulation results show that such embedded method resolves well the contradiction of imperceptibility and robustness. Based on good correlation properties of chaotic sequences, we design a detection method, which can accurately compute geometric transformation (rotation and translation transformations) parameters. The security analysis, including key space analysis, key sensitivity analysis, cryptanalysis, and the comparison results demonstrate that the proposed watermarking scheme also achieves high security.

## 1 Introduction

Digital watermarking is an important technique for protecting the security of information and it is mainly applied in copyright protection. It has been studied more than twenty years since this concept was first proposed in academia [1]. Many watermarking schemes [2-12] have been proposed, among them, several famous algorithms such as spread spectrum (SS) [2] and quantization index modulation (QIM) [3] are investigated, and the earliest research effort has focused on robustness and embedding distortion. A watermarking system should be robust to various attacks, such as noise, general signal processing and

* Corresponding author. Tel. 0086-10-62282452.

*E-mail address:* shwang@bupt.edu.cn.

geometric distortions, etc. How to improve robustness has been a hot issue. Some of the recent studies such as Ref. [4] generally represent state-of-the-art algorithm in robustness performance.

Meanwhile, the security problem of watermarking has been attracting more and more attention. Kalker firstly defines watermarking security as "the inability by unauthorized users to access the communication channel" [13], i.e., to remove, to read, or to write the hidden the message. Later, Comesaña et al. take the view that the goal of security analysis is to gain secret information [14], e.g. embedding the message and detecting keys. In [15], Cayre et al. point out that the security analysis should be based on Kerckhoff's principle [16] that means adversaries know the embedding/detecting algorithms except for a secret key. According to Diffie-Hellman's terminology, they group the security attacks of watermarking systems as (i) Watermark only attack (WOA). (ii) Known message attack (KMA). (iii) Known original attack (KOA). Under WOA scenario, Cayre et al. classify the security levels of watermarking schemes: insecurity, key security, subspace security, and stego-security [17].

In addition, some scholars make quantitative and qualitative analysis on the security of some algorithms, such as SS algorithm [2] and QIM algorithm [3]. Some of watermarking systems are studied in real number field, and the secret keys are real numbers or vectors. On the basis of symmetric cryptography, Bas et al. put forward the concept of the equivalent keys to measure watermarking security [18]. In [3], Cayre et al. establish a general theoretical framework in information theory and analyze SS algorithm by Shannon's mutual information and Fisher's information matrix. Similarly, Pérez-Freire et al. quantitatively analyze the security of the dither modulation QIM (DM-QIM) algorithm under KMA [19] and WOA [20] scenarios. Their investigations show that the security of both SS and DM-QIM are still very low, although SS and QIM algorithms are generally agreed to have good robustness compared with other algorithms. In that case, it is of significant importance to improve the security of the above algorithms.

Chaotic systems have received great attentions in the past decade due to its inherent properties such as non-periodicity, wide spectrum, sensitive to initial conditions and system parameters, and randomness property. Several chaos-based watermarking algorithms are proposed [5-12], where the watermark is embedded in the spatial domain [5-7], or in the frequency domain [8-12]. However, most of the algorithms are still easily analyzed from the cryptographic point of view. For example, in Refs. [9-11], permutation and encryption techniques are applied to protect the watermark information. We find that they are insecure under Kerckhoff's principle, because we can easily access the embedding positions of the watermark, and then remove the watermark, or replace a new encrypted watermark, or decrypt the watermark. In Refs. [5-7], the chaotic sequences are used to determine the embedding positions of the watermark and the embedding positions depend only on the key. As long as we get enough original images and the watermarked images, we can obtain all the embedded positions, i.e., equivalent keys.

In this paper, we attempt to address these security issues. The core idea of the proposed scheme protects not only the watermark information but also the embedding positions. First, an original watermark is encrypted, and then parts of the embedded positions are randomly chosen to embed the encrypted watermark bits. In particular, this process is not only related to the key, but also to the image itself, which will improve the security of the system under the WOA / KOA / KMA scenarios. The main contributions of this paper include: (i) The embedding domain of the proposed scheme is the DCT domain of the scrambled image rather than the DCT domain of the original image. Such improvements try to tackle the contradiction between imperceptibility and robustness; (ii) To resist rotation and translation attacks, a new method based on correlation properties of chaotic sequences is designed, which can accurately estimate geometric transformation parameters; (iii) To resist the cryptanalysis under the WOA / KOA / KMA scenarios, we propose a sub key generation mechanism based on the singular value decomposition and hash function, where sub keys depend on not only the main key but also the feature codes of the image.

The rest of the paper is structured as follows. Section 2 introduces the basic background and theory. Section 3 presents the proposed watermarking scheme in detail. The performance evaluation and security analysis are provided in Section 4 and Sections 5. Finally, a conclusion is drawn in Section 6.

## 2 Preliminaries

### 2.1 Singular value decomposition

Singular value decomposition (SVD) is an orthogonal transform and it is defined as

$$A = USV^T$$

$$= \begin{bmatrix} u_{1,1} & \cdots & u_{1,n} \\ u_{2,1} & \cdots & u_{2,n} \\ & \vdots & \\ u_{n,1} & \cdots & u_{n,n} \end{bmatrix} \begin{bmatrix} \sigma_1 & 0 & 0 & 0 \\ 0 & \sigma_2 & 0 & 0 \\ & & \ddots & \\ 0 & 0 & 0 & \sigma_n \\ \vdots & \vdots & & \\ 0 & 0 & 0 & 0 \end{bmatrix} \begin{bmatrix} v_{1,1} & \cdots & v_{1,m} \\ v_{2,1} & \cdots & v_{2,m} \\ & \vdots & \\ v_{m,1} & \cdots & v_{m,m} \end{bmatrix}^T \quad (1)$$

$$= \sum_{i=1}^{n} \sigma_i u_i v_i^T,$$

where $A$ is an original matrix of size $n \times m$ (suppose $n < m$), and $U$, $V$ are two orthogonal matrices with sizes of $n \times n$, $m \times m$, respectively. $S$ is a singular value matrix, the singular values of which satisfy $\sigma_1 \geq \sigma_2 \geq \ldots \geq \sigma_r > \sigma_{r+1} = \ldots = \sigma_n = 0$.

SVD transformation has two good properties: first, singular values of an image are much less affected after the common image processing, such as filtering, compression. Second, singular values contain intrinsic algebraic image properties. Basing on the two properties, we design the robust feature codes of the image in Section 3.1.

## 2.2 Logistic map

Logistic map is usually used to generate chaotic sequences, and is written as

$$x_i = \mu x_{i-1}(1 - x_{i-1}), \mu \in [3.57, 4], x_0 \in [0,1], \quad (2)$$

where $\mu$ and $x_0$ are the parameter and the initial state value. The digitized sequence $\{k_i\}$ from Logistic map can be generated

$$k_i = \lfloor x_i \times 10^{14} \rfloor \bmod R, \ i = 0,1,2,\ldots,L-1, \ k_i \in \{0,1,2,\ldots,R-1\}, \quad (3)$$

where the operation of $\lfloor x \rfloor$ denotes the largest integer not larger than $x$.

When $R = 2$ in Eq. (3), a binary sequence $\{b_i\}$ can be produced as follows

$$b_i = (-1)^{k_i}, \ i = 0,1,2,\ldots,L-1, \ b_i \in \{-1,1\}. \quad (4)$$

Here we introduce auto-correlation (AC) and cross-correlation (CC) coefficients to evaluate the statistical properties of binary sequences. In [21], AC of $\{b_i\}$ and CC of $\{b_i\}$ and $\{b_i'\}$ are defined as

$$AC(\tau) = \frac{1}{L}\sum_{i=0}^{L-1} b_i b_{i+|\tau|} \tag{5a}$$

$$CC(\tau) = \frac{1}{L}\sum_{i=0}^{L-1} b_i b'_{i+|\tau|} \tag{5b}$$

where $\tau = 0, \pm 1, \pm 2, \ldots$. The two binary sequences $\{b_i\}$ and $\{b'_i\}$ with different initial values are shown in Fig.1 (a) and (b). The auto-correlation coefficients of $\{b_i\}$ and the cross-correlation coefficients of $\{b_i\}$ and $\{b'_i\}$ are shown in Fig.1 (c) and (d), where $L = 10^3$ in Eq. (3).

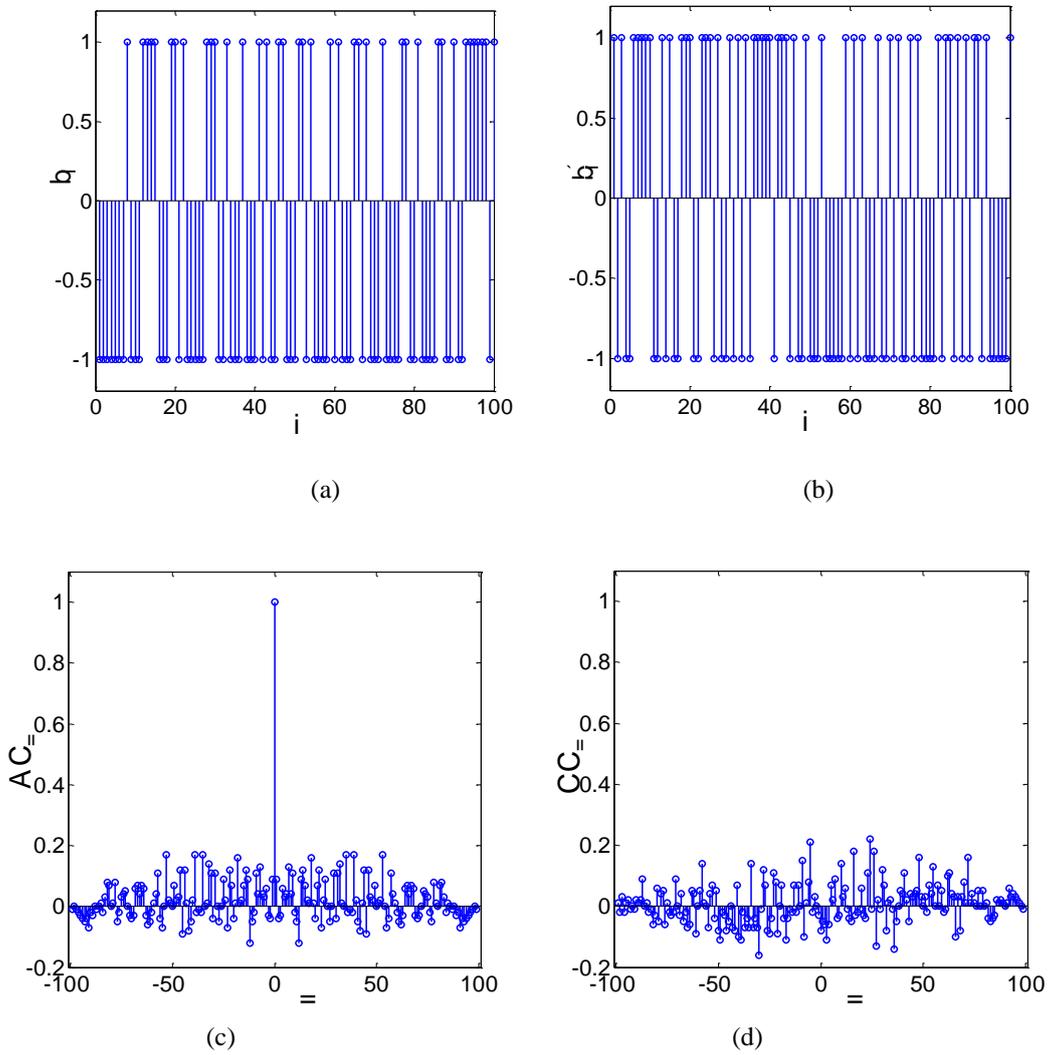

**Fig.1** The binary sequences $\{b_i\}$ and $\{b'_i\}$ with different initial values in (a) and (b). (c) The auto-correlation coefficients of $\{b_i\}$. (d) The cross-correlation coefficients of $\{b_i\}$ and $\{b'_i\}$.

# 3 Proposed watermarking scheme

A complete digital watermarking system should consist of two parts, a transmitter and a receiver. The former is to embed watermark, and the latter to extract watermark. In our proposed scheme, the transmitter's architecture is illustrated in Fig. 2, which is composed of three processes: sub key generation (SKG), embedding process Ⅰ and embedding process Ⅱ. In SKG sub keys are generated and depended on an original image and a key $Key1$. In the embedding process Ⅰ a Logo watermark is embedded in an original image. Further, in the embedding process Ⅱ, an oriented watermark, i.e., a binary chaotic sequence, is embedded in watermarked image I. In the following sections, we describe each process in detail, and briefly introduce extraction process.

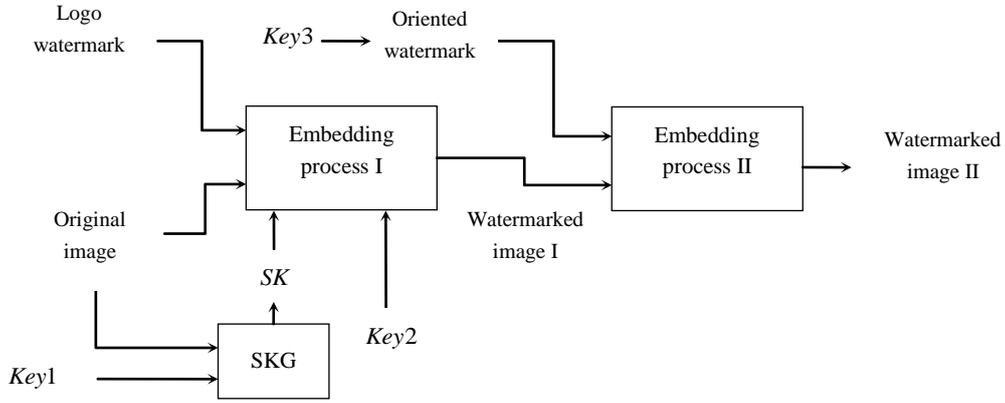

**Fig.2** Architecture of the proposed watermarking scheme

## 3.1 Sub key generation

The framework of sub key generation is shown in Fig. 3. The goal of SKG is to produce different sub keys for different original images at the same main key. To better understand the mechanism, we explain it from two aspects of the transmitter and the receiver.

At the transmitter side, first extract the feature codes of an original image $OI$ by the form

$$FC = Extraction(OI). \qquad (6)$$

Then mix the feature codes $FC$ and the main key $Key1$ by the form

$$TM = Mixing(Key1, FC). \qquad (7)$$

In this paper, we define $TM = Key1 \| FC$, where the symbol $\|$ stands for a concatenation operation. Finally, calculate the hash value of $TM$ and generate a sub key $SK$ by the following form,

$$SK = Hash(TM). \qquad (8)$$

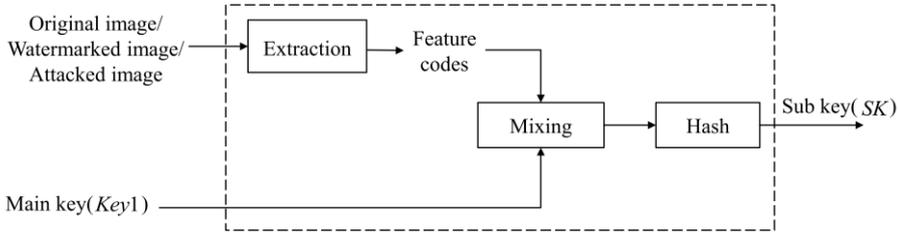

**Fig.3** Sub key generation for watermarking

To extract watermark, the receiver must have the same $SK$. For non-blind watermarking schemes (e.g., SS scheme), the receiver can generate the same $SK$, but for blind watermarking schemes (e.g., DM-QIM scheme), the original image is unavailable. Considering blind watermarking schemes, a feature code extraction method should meet two design requirements:

(1) Robustness: The feature codes of the original image, the watermarked image and the watermarked image being attacked must be identical, i.e., the feature codes should be robust to a few common attacks.

(2) Distinctiveness: the feature codes should be different for different images. In other words, the contents' feature codes are unique.

To meet the above requirements, a simple but effective extraction method based on SVD is presented as follows:

Step 1. First divide $OI$ into $l$ blocks of size $(m+n) \times (m+n)$, where $l = \left\lfloor \dfrac{M}{m+n} \right\rfloor \times \left\lfloor \dfrac{N}{m+n} \right\rfloor$, $m \geq n$. Further divide each block into two sub blocks shown in Fig.4, the sizes of which are $n \times n$. Thus, there exist sub blocks $OI_i$ in the image, $i = 1, 2, ..., 2l$.

Step 2. Calculate the singular value $S_i$ of sub blocks $OI_i$ by Eq. (1). Then obtain the maximum singular value $s_{mi}$ for $S_i$.

Step 3. Compare any two different singular values $s_{mi}$ and $s_{mj}$ to obtain a feature codes sequence $\{FC_k\}$ according to the following rule.

$$FC_k = \begin{cases} 1, & s_{mi} > s_{mj} \\ 0, & s_{mi} \leq s_{mj} \end{cases} \quad i=1,2,...,2l, \ j=i+1,i+2,...,2l. \quad (9)$$

where $k = 1, 2, ..., l(2l-1)$.

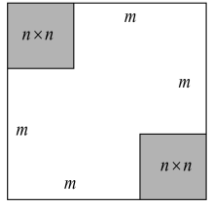

**Fig. 4** Each block of $(m+n) \times (m+n)$ is divided into two sub-blocks, the sizes of which are $n \times n$.

### 3.2 Embedding process I

The embedding procedure I is depicted in Figs. 5. A binary logo watermark $LW$ of size $H \times W$ is embedded in the original image $OI$ of size $M \times N$ ($H \times W < M \times N$), and a watermarked image I is produced. The embedding steps are described as follows:

Step 1. Generate a chaotic sequence $x$ and further get an index sequence $\alpha$. After iterating Eq. (2) with $Key2$ ($\mu$ and $x_0$) for 500 times, we continue to iterate it until a chaotic sequence $x$ of length $M \times N$ is obtained, where $x = \{x_{501}, x_{502}, ..., x_{501+M \times N}\}$. Sort the chaotic sequence $x$ in ascending order, and obtain a new sequence $x' = \{x'_{\alpha_1}, x'_{\alpha_2}, ..., x'_{\alpha_{M \times N}}\}$. Thus an index sequence $\alpha = \{\alpha_1, \alpha_2, ..., \alpha_{M \times N}\}$ is available.

Step 2. Scramble the original image. The original image is scrambled by

$$SI_j = OI_{a_j}, \ j=1,2,...,M \times N, \ a_j \in \{1,2,...,M \times N\} \quad (10)$$

where $OI_j$ and $SI_j$ represent the $j^{th}$ pixels of the original image $OI$ and the scrambled image $SI$, respectively.

Step 3. The scrambled image is transformed by using DCT.

Step 4. Encrypt the logo watermark. Take the first $H \times W$ elements in the sequence $x$, calculate Eq. (3), and obtain a binary sequence $\{k_i\}, i=1,2,...,H \times W$. The logo watermark is encrypted by

$$w'_j = w_j \oplus k_j, \ j=1,2,...,H \times W \quad (11)$$

where the symbol $\oplus$ denotes bitwise XOR. $w_j$ and $w'_j$ represent the $j^{th}$ elements of the original and the encrypted logo watermarks, respectively.

Step 5. Generate another chaotic sequence $y$ and get another index sequence $\beta$. First, an initial value of Eq. (2) is generated from the sub key $SK$ by the following form

$$y_0 = \frac{\sum_{i=0}^{len} 2^i SK_i}{2^{len} - 1}, \qquad (12)$$

where $SK_i$ denotes the $i^{th}$ bit of the sub key $SK$, and $len$ the length of $SK$. Then, similarly as Step 1, with $\mu = 3.99999999999999$, a new chaotic sequence $y$ of length $M \times N$ is obtained. Similarly as the Step 1, sort the sequence $y$ to obtain another index sequence $\beta = \{\beta_1, \beta_2, ..., \beta_{M \times N}\}$.

Step 6. The encrypted logo watermark is embedded in the DCT coefficients of the scrambled image. Take the first $H \times W$ elements of the index sequence $\beta$ to determine the embedding positions. For example, the first watermark bit is embedded in the $(\beta_1)^{th}$ coefficient, the second watermark bit is embedded in the $(\beta_2)^{th}$ coefficient, etc. To achieve blind extraction, QIM method [3] is used as follow

$$Q(c_i) = \begin{cases} Q_0(c_i) = \text{round}\left(\frac{c_i}{\Delta}\right) \times \Delta, & \text{if } w'_i = 0, \\ Q_1(c_i) = \text{round}\left(\frac{c_i - 0.5 \times \Delta}{\Delta}\right) \times \Delta + 0.5 \times \Delta, & \text{if } w'_i = 1, \end{cases} \qquad (13)$$

here $c_i$ is the $i^{th}$ of the selected DCT coefficient, $\Delta$ denotes the quantization step size, and the operation of round$(x)$ an integer to the nearest $x$.

Step 7. Obtain the watermarked image I. The embedded image is transformed by using IDCT and inverse permutation, and the watermarked image I is obtained.

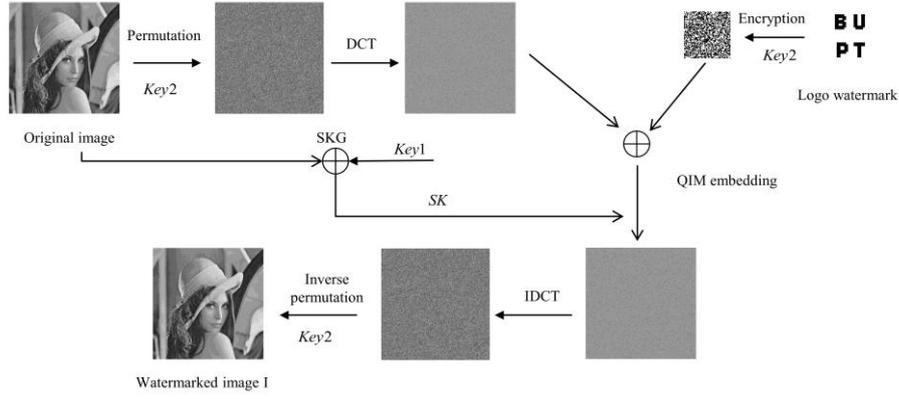

**Fig.5** Embedding process I: a logo watermark is embedded

### 3.3 Embedding process II

To resist against geometric attacks, a binary chaotic sequence called an oriented watermark is embedded twice after the logo watermark is hidden. The embedding procedure II is illustrated in Fig. 6, and includes the following two steps:

Step 1. Generate an oriented watermark $OW$. Similarly as the Step 1 of Section 3.2, a new chaotic sequence $z$ is obtained with $Key3$ ($\mu$ and $z_0$), and the sequence length $len_{ow} \leq 0.5 \times \min(M, N)$. Then applying Eq. (3) generates a binary sequence as an oriented watermark $OW$.

Step 2. Construct an orientation template and embed the oriented watermark $OW$. Choosing the central position $(\lfloor 0.5 \times M \rfloor, \lfloor 0.5 \times N \rfloor)$ in the watermarked image I as a starting point (i.e., the axis origin), we randomly fix two symmetrical directions, $90° - \theta$ and $90° + \theta$, $\theta \in (0, 90°)$, as an orientation template. Note that we do not need to record the two directions, but the two directions are symmetrical and their synthetic direction is $90°$, further we define it as the marked direction. Then the oriented watermark $OW$ is embedded in the two directions by using QIM method (Eq. (13)) with embedding intensity $\Delta'$.

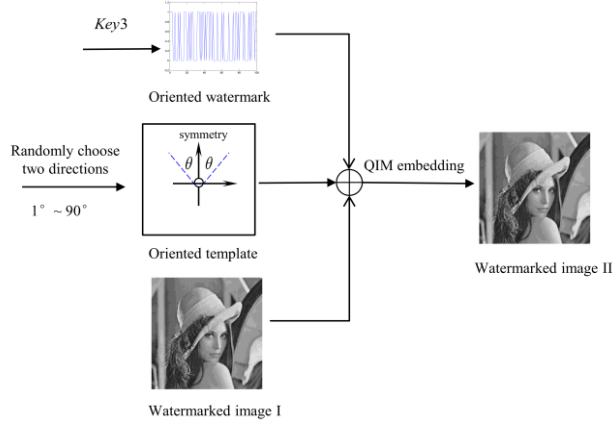

**Fig.6** Embedding process II: the same oriented watermark is embedded twice in the two directions.

**Remark 1:** By using QIM method, the logo watermark and the oriented watermark are embedded in the frequency and the spatial domains, respectively. The latter's embedding does not affect the former's when the embedding intensities satisfy the following form $\Delta \geq 1.5\Delta'$.

### 3.4 Watermark detection and extraction process

The process is shown in Fig.7, and it can be further divided into two sub processes: the oriented watermark detection process and the logo watermark extraction process. The purpose of the former is to determine the geometric transformation parameters of the attacked image and to rectify and reshape the image based on these parameters, and the purpose of the latter is to extract the logo watermark from the reshaped image.

The oriented watermark detection process is described as follows.

Step 1. Similarly as step 1 in Section 3.3, construct an original oriented watermark $OW$ with $Key3$.

Step 2. Detect the positions and orientations of the two oriented watermarks in the received image. Choose the central position of the received image as a starting point (i.e., the axis origin), extract the possible binary sequences $\overline{OW}_\varphi$ by using Eq. (14) at the direction of $\varphi$, where $\varphi = 1°, 2°, 3°, ..., 359°, 360°$.

$$\overline{ow}_{\varphi,i} = \arg\min_k \left\{ \left\| \overline{c}_{\varphi,i} - Q_k \right\| \right\}, k \in \{0,1\}, \tag{14}$$

where $\overline{c}_{\varphi,i}$ and $\overline{ow}_{\varphi,i}$ represent the $i^{th}$ of the pixel and corresponding watermark bit, $Q_0$ and $Q_1$ denote two different quantizers. We calculate the corresponding cross-correlation value $CC_\varphi(0)$ between $\overline{OW}_\varphi$ and $OW$ by using Eq. (5b). We plot the curve of $CC_\varphi(0)$ vs. $\varphi$ and find the two largest peaks, and the corresponding directions are recorded $\delta_0$ and $\delta_1$. Thus rotation angle of the received image can be calculated as the form

$$\Delta\delta = \frac{(\delta_1 + \delta_2)}{2} - 90°. \qquad (15)$$

Step 3. Rectify and reshape the image. According to Eq. (15), we rotate the image counterclockwise by $\Delta\delta$, and obtain a reshaped image.

**Remark 2:** Note that the image may suffer from a translation attack. In this case we can translate images by different distances, and then re-performing correlation detection and calculate parameters. Some examples of correlation detection are given in Section 4.5.

The logo watermarking extraction process is the inverse of the embedding process I, which is briefly described in the following steps.

Step 1. Obtain the sub key $SK$ through the SKG process with $Key1$ and the reshaped image.

Step 2. Apply permutation and DCT transform on the reshaped image, and then extract the encrypted watermark from the DCT coefficients. Note that the embedding positions are determined by $SK$, extract the watermark bits from the coefficients of the positions using the Eq. (14).

Step 3. Decrypt the encrypted watermark with $Key2$.

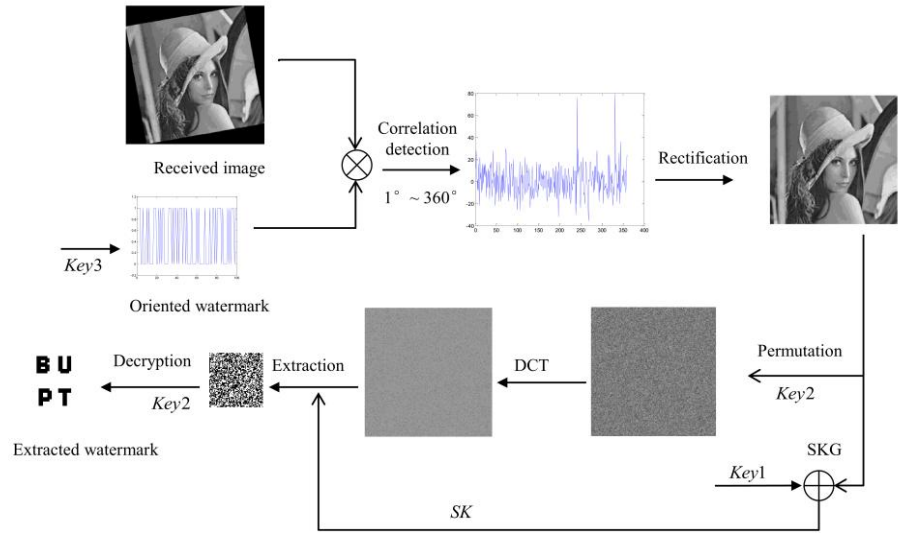

**Fig.7** Procedure of the extracting the logo watermark

# 4 Simulation results and discussions

In this section, simulation experiments are used to verify the validity of the proposed SKG mechanism, then to demonstrate the superiority and effectiveness of watermark embedded in DCT domain of the scrambled image. Finally, we evaluate the performance of the whole watermarking system in both imperceptibility and robustness.

## 4.1 Experimental setup and evaluation metrics

In the following experiments, five USC-SIPI test gray-scale images with size $512 \times 512$ are adopted as the original images, including Lena, Barbara, Baboon, Peppers, and Goldhill. A $64 \times 64$ binary image (BUPT) is chosen as a logo watermark, while two binary pseudo-random sequences of 150-bit are used as an oriented watermark, which is determined by $Key3$, i.e., $\mu = 3.979986502162631$ and $z_0 = 0.64246416698522$. The quantization steps can be used by $\Delta = 36$ and $\Delta' = 18$ in the two embedding processes. In the SKG, we set $m = 100$, $n = 68$, and 18 singular values and 153-bit feature codes $FC$ can be obtained. The simulation is implemented in MATLAB 2014a.

To quantitatively evaluate the performance of the system, we employ various performance metrics, such as peak signal to noise ratio (PSNR), structural similarity (SSIM) [22], normalized cross correlation (NC) and bit error rate (BER). PSNR and SSIM are used to evaluate the quality of the watermarked images or the attacked versions. PSNR is defined by the following forms

$$\text{PSNR} = 10\log_{10}\frac{255^2}{\text{MSE}},$$
$$\text{MSE} = \frac{1}{M \times N}\sum_{i=1}^{i=M}\sum_{j=1}^{j=N}\left(WI_{i,j} - OI_{i,j}\right)^2, \quad (16)$$

where $OI_{i,j}$ and $WI_{i,j}$ denote the pixels of the original and the watermarked images.

SSIM is defined by the form

$$\text{SSIM} = \left(\frac{2\mu_X\mu_Y + C_1}{\mu_X^2 + \mu_Y^2 + C_1}\right)^{\lambda 1}\left(\frac{2\sigma_X\sigma_Y + C_2}{\sigma_X^2 + \sigma_Y^2 + C_2}\right)^{\lambda 2}\left(\frac{2\sigma_{XY} + C_3}{\sigma_X\sigma_Y + C_3}\right)^{\lambda 3}, \quad (17)$$

where $\mu_X$ and $\mu_Y$ are the means of the original image $OI$ and the watermarked image $WI$, respectively. $\sigma_X$ and $\sigma_Y$ are the variances of $OI$ and $WI$, respectively, and $\sigma_{XY}$ is the covariance between $OI$ and

$WI$. $\lambda 1, \lambda 2, \lambda 3, C_1, C_2$ and $C_3$ are constants. SSIM ranges in [0, 1], and the larger the value, the smaller the image distortion.

Meanwhile we adopt NC and BER to measure the accuracy of the extracted watermark. They are defined as

$$\text{NC} = \frac{\sum_{i=1}^{H \times W} W_i W_i'}{\sqrt{\sum_{i=1}^{H \times W} W_i^2 \sum_{i=1}^{H \times W} W_i'^2}}. \tag{18}$$

$$\text{BER} = \frac{\text{num of } (FC_i' \neq FC_i)}{\text{length of } FC}. \tag{19}$$

where $W_i$ and $W_i'$ are the $i^{th}$ bit of the original watermark and the extracted watermark, respectively.

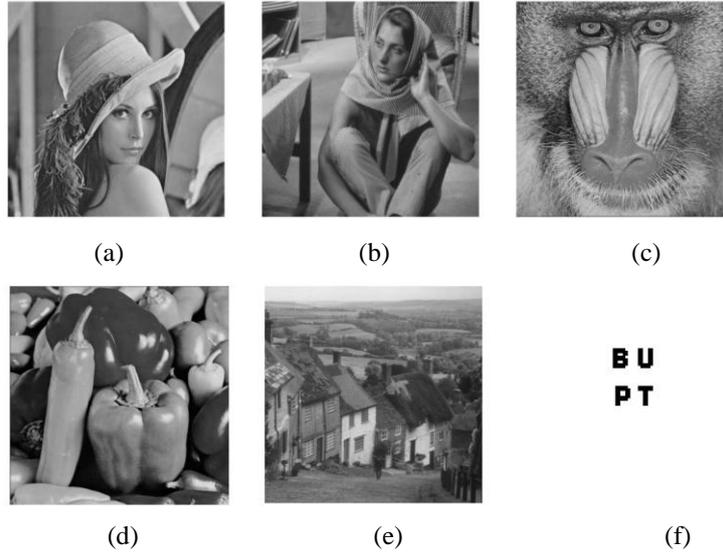

(a)　　　　　　　　(b)　　　　　　　　(c)

(d)　　　　　　　　(e)　　　　　　　　(f)

**Fig. 8** Five test gray-scale images of size $512 \times 512$ and a logo watermark of size $64 \times 64$. (a) –(e)Lena, Barbara, Baboon, Peppers and Goldhill. (f) A logo watermark.

### 4.2 Effectiveness analysis of the SKG

*4.2.1 Feature codes and sub keys for different images*

In this experiment, we extract the feature codes of different images and obtain corresponding sub keys. For different security levels, the lengths of the main key can be chosen 128-bit, 256-bit and 512-bit. In simulations, we take a 128-bits main key $Key1$ and SHA3-256 algorithm with block length 1088 bits. We concatenate the feature codes $FC$ with the main key $Key1$, subsequently pad zeroes such that 1088-bit length the message is obtained, i.e., $TM = Key1 \| FC00...0$. After SHA3-256 algorithm, 256-bit sub key is generated.

Table 1 shows the feature codes of five images and the corresponding sub keys. In simulations we arbitrarily take a $Key1$ with hexadecimal format $Key1$=bae63457983b9e7d052f5867dee30024. The simulation results show that the five different images have distinguishing feature codes, and the obtained sub keys have good randomness due to hash functions.

**Table 1** Different feature codes and the corresponding sub keys for different images

| Images | Feature codes | Sub keys |
| --- | --- | --- |
| Lena | 40800102de8913ffed9a006bdf8000ff00f64d8 | 71cc0b895af2a238faee01a2e2db41dfafb583657bfabcdd87b5266863dd2857 |
| Barbara | f841c3020c183f7eff800061d7870fbf7effc14 | 8ffce23b6edf07b01054d08732a33029e03c1361c849c4b5f459e66a273213f8 |
| Baboon | 1eff7fff00890235ebd62048bc00f783078f718 | aa562485d279bd1501601c2c5c8201e895898fb62d1380d7457087827a400b12 |
| Peppers | 40a300cf1b3fffff3a74e9800042bd08fea87a0 | 3a463cbb423b9846f64a0e7b8d7fb5c1188952f8608bb4be8cdabee0b2d0bb98 |
| Goldhill | 8000001830fefc000066cd9b276fdf3e7cff020 | 1388bd151a6186305f8daa447e32184bf0e87e64a5edcbbc0d3aec71acdb9263 |

*4.2.2 Robustness of feature codes*

The feature codes of transmitter and receiver should be identical in the watermarking system, in other words, the feature codes should be robust to image processing or some attacks. Assume the channel is ideal without noise and any attacks. We can compare the features codes of the original and the watermarked images, and find that the features codes of five pairs of images are synchronous. The results are presented in Table 2.

**Table 2** Features codes synchronization of the watermarked and the original images

| Watermarked/ Original images | Lena | Barbara | Baboon | Peppers | Goldhill |
| --- | --- | --- | --- | --- | --- |
| PSNR | 45.57 | 45.50 | 45.64 | 45.67 | 45.55 |
| Features codes synchronization？ | **Yes** | **Yes** | **Yes** | **Yes** | **Yes** |

In fact, the channel is not ideal and there exist noise and malicious attacks. Considering image processing, such as compression, filtering and geometric transformation, we should investigate the features codes of the attacked watermarked image. In particular, we focus on the additive Gaussian white noise (AGWN) and JPEG compression, the results of which are shown in Table 3 and 4, respectively. As can be seen from Table 3, the larger the noise variance, the worse the quality of the watermark image (PSNR). But the features codes are still synchronized, even when variance $\sigma = 20$, PSNR is reduced to $22.09 \, dB$. These results represent that this method to extract features codes is robust to AWGN.

**Table 3** Additive Gaussian white noise with zero mean value against watermarked Lena.

| Variance $\sigma$ | 1 | 2 | 5 | 10 | 16 | 20 |
|---|---|---|---|---|---|---|
| PSNR(dB) | 43.53 | 40.43 | 33.84 | 28.05 | 24.01 | 22.09 |
| Features codes synchronization？ | **Yes** | **Yes** | **Yes** | **Yes** | **Yes** | **Yes** |

In Table 4, we test the different compression factors for JPEG compression, and we find that the features codes are synchronized if compression factor $Q = 10$.

**Table 4** JPEG Compression on watermarked Lena for different factors $Q$

| Factor $Q$ | 100 | 90 | 80 | 50 | 20 | 10 |
|---|---|---|---|---|---|---|
| PSNR(dB) | 45.28 | 39.98 | 38.17 | 35.68 | 32.95 | 30.39 |
| Features codes synchronization？ | **Yes** | **Yes** | **Yes** | **Yes** | **Yes** | **Yes** |

Furthermore, we test some other types of attacks, and the results are shown in Table 5. We find that the feature codes are synchronized under mean filtering (the filter template size is $3 \times 3$), Gaussian low pass filtering attacks, the rotation transformation (the rotation angle is 8°) and translation transformation (the translation distance is 40 pixels). But the feature codes designed by us cannot resist the cropping attack (the cropping ratio is 25%), because they are based on blocks.

**Table 5** Other attacks on watermarked image Lena

| Attacks | Mean filtering (Template [$3 \times 3$]) | Gaussian low pass filtering | Cropping (25%) | Translation (40 pixels) | Rotation (8°) |
|---|---|---|---|---|---|
| PSNR(dB) | 31.90 | 40.26 | _ | _ | _ |
| Features codes synchronization？ | **Yes** | **Yes** | No | **Yes** | **Yes** |

## 4.3 Effectiveness of watermark embedded in the DCT domain of the scrambled image

In the proposed scheme, the watermark is embedded in the DCT domain of the scrambled image that has the following two advantages:

(i) High embedding capacity up to $(M \times N - 1)$ bits. The DCT coefficient distribution of the scrambled image is similar to that of noise, which provides more embedded positions (except the direct component (DC) coefficient). Simulation results are shown in Fig. 9. The DCT coefficients (except DC coefficient) of the scrambled Lena image are distributed in [-300,300] and similarly as that of noise, while the DCT coefficients of the original image (except DC coefficient) are distributed in [-5200, 5200], and there are about $3 \times 10^4$ coefficients close to zero, they cannot be used as an embedded position due to being easily erased by image processing. To further verify that the DCT energy of the scrambled image is not concentrated in the upper right corner (low frequency region), we calculate the energies of the upper right corner of the two DCT coefficient matrices (size $128 \times 128$, but except for the DC energy), and then divide by the total energy of the image (also except for the DC energy), defined as energy ratio. The experimental results are presented in Table 6. The average energy rate of five original images is about 90.53%, while the average energy rate of five scrambled images is about 6.26%, being approximately equal to the theoretical value 6.25% (i.e., $\frac{(128 \times 128 - 1)}{(512 \times 512 - 1)}$).

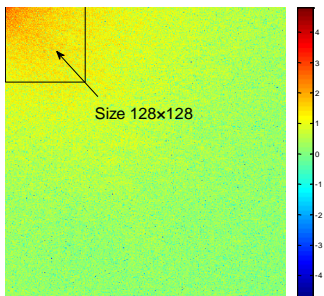

(a)

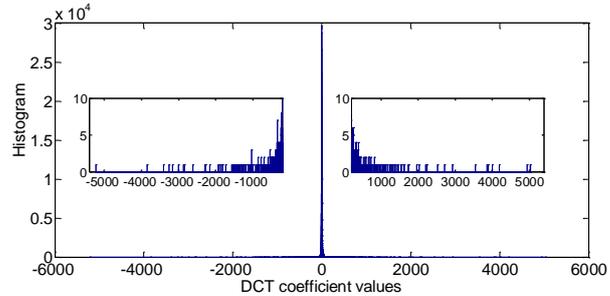

(b)

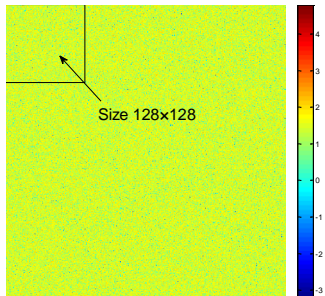

(c)

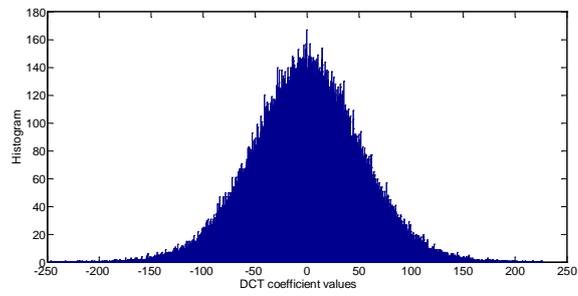

(d)

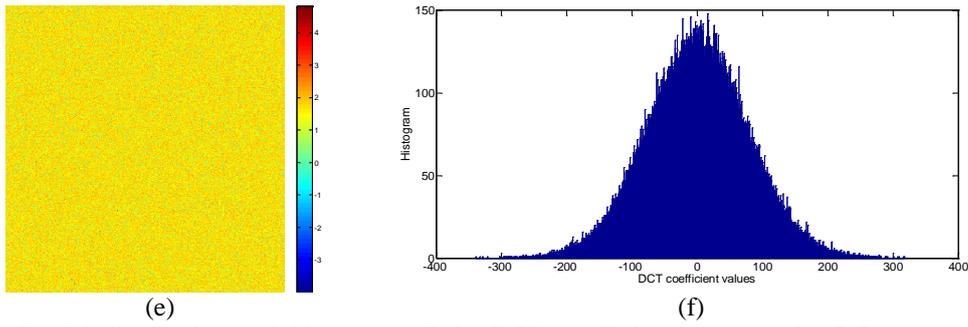

(e) (f)

**Fig.9** Spatial distribution and histogram of the DCT coefficients (except the DC component). (a-b) Original Lena image. (c-d) Scrambled Lena image. (e-f) Noise image.

Table 6 DCT coefficient distribution range and the energy ratios of the upper right corner of the (size $128\times128$, but except for the DC coefficient) of DCT coefficient matrices.

| Images | Attributes | Lena | Barbara | Baboon | Peppers | Goldhill |
| --- | --- | --- | --- | --- | --- | --- |
| Original image | Range | [-5215.5, 5038.0] | [-4948.4, 4771.9] | [-4608.3, 5230.4] | [-6647.2, 7322.6] | [-4306.4, 10355.2] |
| | Energy rate | 97.08% | 88.48% | 73.61% | 97.56% | 95.95% |
| Scrambled image | Range | [-236.9, 233.9] | [-218.7, 230.1] | [-224.1, 187.2] | [-283.1, 296.6] | [-246.7, 225.7] |
| | Energy rate | 6.27% | 6.28% | 6.31% | 6.25% | 6.18% |

(ii) Robust performance is independent of the selected frequency regions. For the different regions of the DCT coefficient matrix of the original image, the robustness performances are different under the JPEG compression: the low frequency (LF) region shows best robustness, then the intermediate frequency (IF) region, and the high frequency (HF) region is worst. However, from the Human Visual System (HVS) point of view, the human's eye is very sensitive to the low frequencies and is not sensitive to the high frequency coefficients. To reach a compromise between robustness and visual invisibility, the watermark is usually embedded in the intermediate frequency region. However, the proposed scheme does not need to consider the embedded frequency region, because all frequency regions provide the same robust performance, especially under JPEG compression. In a sense, the proposed scheme solves the contradiction between imperceptibility and robustness.

Next we take the simulation results and prove the above conclusion (ii). For the sake of convenience, we define the upper left corner of the DCT coefficients matrix (size $512\times512$) as the coordinate origin (1,1). Then we define three rectangular regions of size $64\times64$, LF: from coordinates (11, 11) to (74, 74);

IF: from coordinates (221, 221) to (284, 284); HF: from coordinates (431, 431) to (494, 494). These three regions, LF, IF and HF, represent low, intermediate and high frequency regions for embedding watermark, respectively. To better verify the conclusion, we conducted two sets of comparative experiments. In the first set of experiments, we use the proposed scheme and embed the logo watermark in the LF, IF and HF regions of the five images, which produce 15 different watermarked images. Then we test these images under JPEG compression with different quality factors and extract the corresponding watermarks. Finally calculate the NC and BER values for these watermarks and calculate their mean values by embedding region. In another set of experiments, the only difference is that the scheme doesn't include permutation process, i.e., the three embedding regions are chosen from the DCT coefficients matrix of the original image. The results of the two sets of experiments are shown in Figs.10 (a-b). Two conclusions can be drawn from the figure: (i) For the original image (OI, i.e., without permutation process), the different regions of the DCT coefficient matrix have different robustness performances under JPEG compression. LF has the strongest robustness and can resist the compression at quality factor $Q \geq 25$ (Here we set the threshold $NC \geq 0.8$ or $BER \leq 0.3$), while HF has the weak robustness and can only resist the compression at $Q \geq 90$. IF has the moderate robustness and can resist the compression at $Q \geq 80$. (ii) For the scrambled image (SI), the robustness performances of the different regions at $Q \geq 70$ are almost the same.

We present experimental results under AGWN attacks in Figs.10 (c-d). It can be seen that all the curves are coincident approximately, because the noise generated is independent of the images and its frequency distribution is evenly distributed.

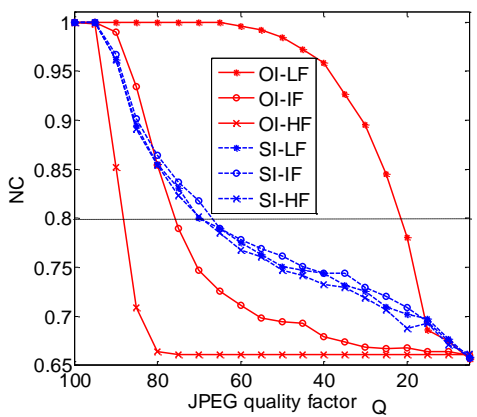 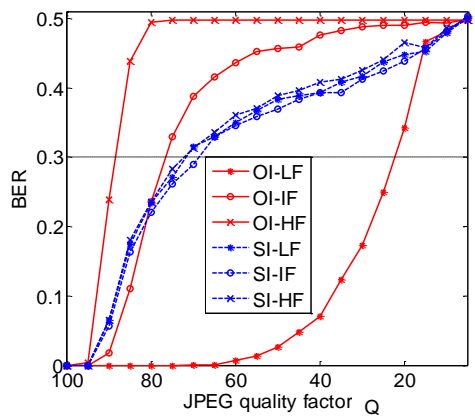

(a) (b)

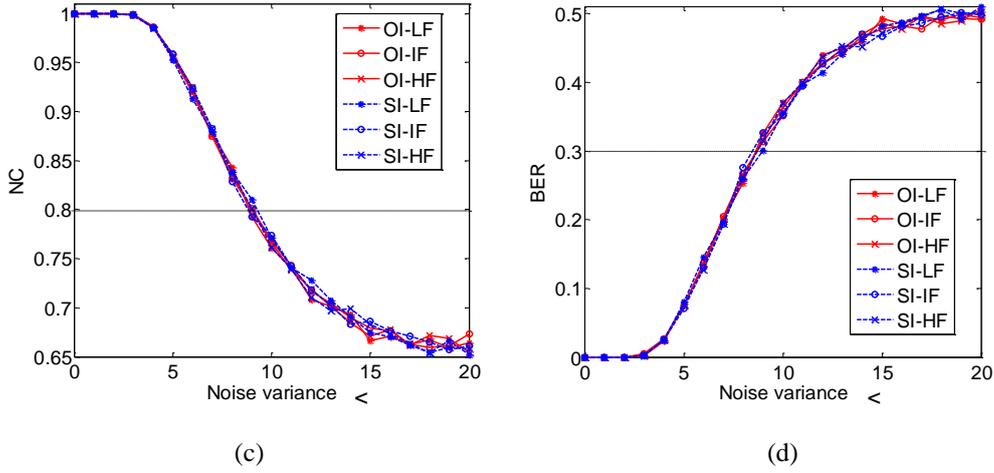

(c)    (d)

**Fig. 10** Comparisons of robustness performance in different embedding regions by using two methods. SI and OI stand for our proposed method and the method without permutation process, respectively. (a-b) JPEG compressions. (c-d) AGWN attacks. The blue dotted lines indicate the proposed scheme, and the red solid lines indicate that the proposed scheme without permutation process. LF, IF and HF represent low, intermediate and high frequency embedding regions, respectively.

### 4.4 Imperceptibility evaluation of the watermarking system

There are two watermark embedding procedures in our scheme. First, a logo watermark with 4096 bits capacity is embedded, and then an oriented watermark with 150 bits capacity is embedded twice. In this way, the watermarked images I and the watermarked images II will be output.

Table 7 presents the values of PSNR and SSIM for the 5 test images in two embedding processes. From Table 7, we can observe that $PSNR \geq 45.61 \text{(dB)}$ and $SSIM \geq 0.99$ for the watermarked images I and there are no declines for the watermarked image II. The results indicate that the proposed scheme has good image perceptual quality.

**Table 7** PSNR and SSIM of the watermarked and the original images

| Watermarked images | | Lena | Barbara | Baboon | Peppers | Goldhill |
|---|---|---|---|---|---|---|
| Embedding I | PSNR(dB) | 45.68 | 45.61 | 45.73 | 45.79 | 45.64 |
| | SSIM | 0.99 | 1.00 | 0.99 | 0.99 | 0.99 |
| Embedding II | PSNR(dB) | 45.57 | 45.50 | 45.64 | 45.67 | 45.55 |

|  | SSIM | 0.99 | 0.99 | 0.99 | 0.99 | 0.99 |

### 4.5 Robustness evaluation of the watermarking system

To evaluate watermark robustness to common image processing operations, here we consider AGWN with different intensity, the JPEG compression with different quality factors, rotation and translation transformations, and filtering. The detailed simulation results are listed in Table 8.

**Table 8** Watermark robustness to common image processing operations.

| Attacks | PSNR | Extracted logo | BER | NC |
|---|---|---|---|---|
| AGWN $\sigma = 1$ | 43.53 | 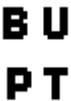 | 0 | 1.00 |
| AGWN $\sigma = 2$ | 40.43 | 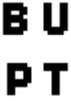 | 0 | 1.00 |
| AGWN $\sigma = 5$ | 33.84 | 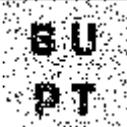 | 0.07 | 0.96 |
| AGWN $\sigma = 8$ | 29.94 | 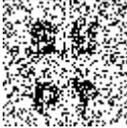 | 0.26 | 0.84 |
| JPEG $Q = 100$ | 45.28 | 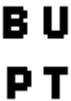 | 0 | 1.00 |
| JPEG $Q = 95$ | 41.74 | 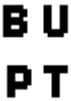 | 0 | 1.00 |
| JPEG $Q = 90$ | 39.97 | 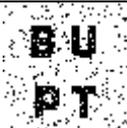 | 0.06 | 0.96 |
| JPEG $Q=80$ | 38.16 | 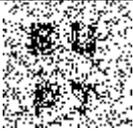 | 0.23 | 0.86 |
| JPEG $Q=75$ | 37.55 | 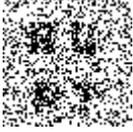 | 0.27 | 0.83 |
| Mean filtering [3×3] | 31.90 | 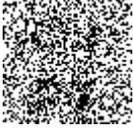 | 0.38 | 0.75 |
| Gaussian low pass filtering | 40.26 | 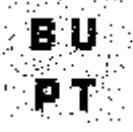 | 0.03 | 0.98 |

It can be seen that the proposed scheme is robust to AGWN with intensity $\sigma \leq 8$, the watermark can be correctly extracted with $BER=0.26$ and $NC=0.84$. The scheme is also robust to the JPEG compression when $Q \leq 80$, $BER$ and $NC$ of the extracted watermark are 0.23 and 0.86, respectively. For filtering operations, it can resist Gaussian low-pass filtering, but cannot resist mean filtering.

To verify the proposed algorithm can resist rotation and translation transformations, we do the following three experiments on Lena image.

**E1:** Detection of the oriented watermark without geometric attack. In the embedding process II, the oriented watermark is embedded in $30°$ and $150°$ directions, respectively (Fig.12 (a)). The receiver extracts the possible binary sequences at all directions, then performs correlation detection with the original oriented watermark. The detection results are shown in Fig.12 (b).

**E2:** Detection of the oriented watermark under a rotation attack. The watermarked image is rotated $8°$ counterclockwise (Fig.12 (c)). We detect similarly as **E1**, and the results are presented in Fig.12 (d).

**E3:** Detection of the oriented watermark under a translation attack. For simplicity, we assume that the watermarked image is shifted 40 pixels in $30°$ clockwise direction. (Fig.12 (e)). The receiver extracts the possible binary sequences at different shifting (in $30°$ clockwise direction) and performs correlation detection with the original oriented watermark. The results are shown in Fig.12(f).

As can be seen from Figs.12 (b) and (d), two peaks of the correlation sequence are at $30°$ and $150°$ (in (b)), at $38°$ and $158°$ (in (d)), and the corresponding marked direction are $90°$ ($(30°+150°)/2$) and $98°$ ($(38°+158°)/2$). The marked direction $90°$ indicates that rotation and translation attacks have not been found in Fig. 12 (a), while the marked direction $98°$ indicates to rotate $8°$ counterclockwise in Fig.12 (c). In Fig.12 (f), obviously we conclude that the received image in Fig.12 (e) is shifted 40 pixels.

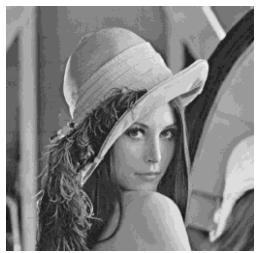
(a)
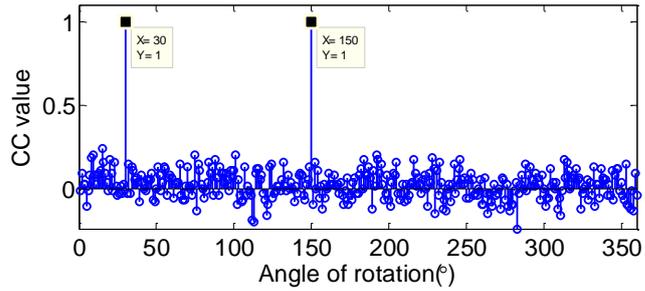
(b)

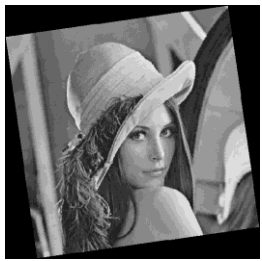
(c)
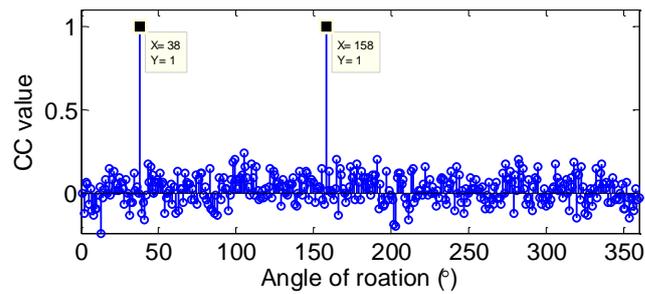
(d)

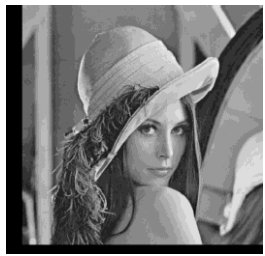
(e)
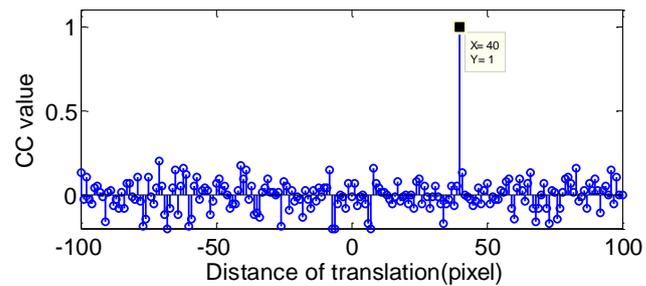
(f)

**Fig. 12** Received images and corresponding detection results. (a-b) Without geometric attack. (c-d) Rotation attack with $8°$ counterclockwise. (e-f) Translation attack with 40 pixels in $30°$ clockwise direction.

# 5 Security analysis and comparisons

## 5.1 Key space

The key space should be large enough to resist against brute-force attacks. The proposed system uses three types of keys, $Key1$, $Key2$ and $Key3$. As mentioned in 4.2.1, $Key1$ is a binary string whose length can be 128, 256 and 512 bits. $Key2$ and $Key3$ are made up of two floating point numbers $(\mu, x_0)$, where $x_0 \in [0,1]$ and $\mu \in [3.57, 4]$. In our simulation, 64-bit double precision is adopted. According to the IEEE floating-point standard [24], the precision of the computer is about $10^{-15}$. Thus, the key space of $Key2$ or $Key3$ are approximately $10^{15} \times (4.3 \times 10^{14}) = 4.3 \times 10^{29} \approx 2^{98}$. With three modes of $Key1$, the total key spaces are $2^{324}$, $2^{452}$ and $2^{708}$, respectively.

## 5.2 Key sensitivity

We first check the sensitivity of $Key2$. Given a $Key2$ ($\mu, x_0$) and a wrong $Key2$ ($\bar{\mu} = \mu + 10^{-15}$, $\bar{x}_0 = x_0 + 10^{-15}$), we calculate corresponding scrambled images and the differential image shown in Figs. 13(a-c). The differential image shows the satisfactory sensitivity of $Key2$ because the behaviour of chaotic system is sensitive to the initial conditions and the parameters. To check the sensitivity of $Key1$, we only change 1 bit of a $Key1$, i.e., bae63457983b9e7d052f5867dee30025, to detect the watermark and the results are shown in Figs. 13(d-f). As the same as $Key2$, $Key1$ has a satisfactory sensitivity.

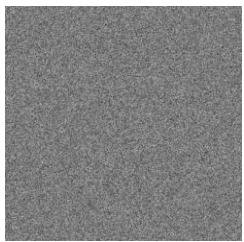  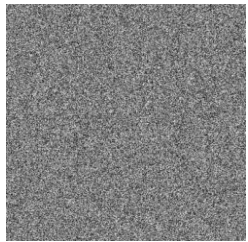  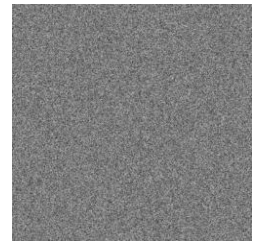

(a)  (b)  (c)

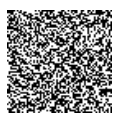  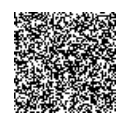

(d) (e) (f)

**Fig. 13** Results of key sensitivity analysis. (a-b) Scrambled image with right and wrong $Key2$, respectively. (c) Differential image of (a) and (b). (d-e) Extracted watermark with right and wrong $Key1$, respectively. (f) Differential image of (d) and (e).

### 5.3 Cryptanalysis under WOA / KOA / KMA scenarios

The goal of cryptanalysis of watermarking systems is to obtain all or partial keys, or equivalent keys. After obtaining the keys, the attacker can do the following thing [24]: unauthorized embedding, unauthorized decoding and unauthorized removal. The security of the proposed scheme mainly depends on $Key1$ and $Key2$, where $Key1$ and the feature codes of the image together determine the embedding position, and $Key2$ is used to scramble the original image and encrypt the logo watermark.

Now we analyze the whole watermarking system under different attack scenarios. In the WOA scenario, the attacker obtains only a series of watermarked images. Assume that he/she made some statistical analysis in the spatial domain of the watermarked images, but it is difficult to analyze in this case. Because the watermark bit is embedded in the different DCT coefficients for different original images, and after IDCT transform, the watermark bit will spread to the whole space domain.

In the KOA scenario, an attacker can obtain some watermarked images and the corresponding original image. An attacker may best use a differential attack method, i.e., he/she calculates the difference between the watermarked image and the original image to get some information about the keys. But the differences are not the identical, because the selected embedding coefficient depends on not only $Key1$ but also the feature codes of different image. So it is difficult to recover the keys.

In the KMA scenario, an attacker collect several watermarked image and watermarks pairs. In this case, he/she may establishes mapping relationship from different watermarks to different watermarked images. It is also difficult to break the system, because as long as the image contents or the feature codes are changed, the embedding positions of the watermark will be changed.

To further demonstrate the high security of the system, we assume that some of the keys are leaked, i.e., or permutation matrix is known to the attacker. The attacker will performs the permutation and DCT transform, then he can guess the embedding position and the embedding order of the watermark. The probability of success can be calculated as the following form

$$P_{\text{sucess\_1}} = \frac{1}{C_{512\times512}^{64\times64} A_{64\times64}^{64\times64}} \approx 6.09 \times 10^{-22185}, \qquad (20)$$

This shows that it is almost impossible to extract the correct encrypted watermark without $Key1$. Furthermore, suppose the attacker already has a part of the sub keys, he/she will predict the main key or the other sub key with less cost. The difficulty of breaking SKG is equivalent to break hash functions. Based on the above analysis, the security of the proposed watermarking system with SKG is greatly enhanced by one-way hash functions and one-time pad mechanism.

**5.4 Comparisons with other schemes**

In this section, we compare our scheme with two representative watermarking schemes, Liu's scheme [9] and Behina's scheme [5]. They are both blind watermarking algorithm with the secret keys, and simple description of the two algorithms are given here.

**Liu's scheme** [9]. First, an original watermark is scrambled based on Logistic maps. Then a scrambled watermark is embedded in the sub-band LL2 of DWT coefficients. Note that the parameters of the Logistic system are the secret keys of the watermarking system.

**Behina's scheme** [5]. Three chaotic sequences are generated based on a coupling chaotic system, two of which are used to determine which pixels of the original image and the other to determine which bit-plane in the pixels to embed watermark using the substitution method. The parameters of coupling chaotic system are the secret keys of the watermarking system.

The comparison results are listed in Table 9. For Liu's scheme, based on Kerckhoff's principle, an attacker is very easy to access the positions of embedding watermark bits. Under WOA or KOA scenarios, He/she can do the following illegal operation, for example, remove the watermark by modifying the DWT coefficients of the watermarked image, extract the encrypted watermark by using same rules, and then embed the extracted encrypted watermarks into other images (i.e., forge the encrypted watermark into a new image). Under the KMA attack, the attacker maybe obtains some original watermarks. By Li's analytical methods [26], only $\log_2(64\times64)=12$ (where $64\times64$ denotes the size of the watermark used in [9]) different known watermarks are needed to acquire the equivalent key, i.e., the permutation matrix.

For Behina's scheme, we know that the embedding positions in the spatial domain are completely dependent on the keys, so the set of all embedding positions can be regarded as the equivalent key of the system. When a certain pixel of the original image is different from that of the watermarked image, it determines an embedded position. The probability of successful confirmation can be estimated by the following equation

$$P_{\text{sucess\_2}} = 1 - \left(\frac{1}{2}\right)^{num}. \tag{21}$$

When $num = 5$ and $num = 10$, approximately 96.9% and 99.9% of the embedding positions were revealed, respectively. Through these embedding positions and extracted watermark, attackers can erase the watermark and move the watermark into a new image.

Basing on the above analysis, we know that Liu's scheme is insecure under WOA, KOA and KMA scenarios, and Behina's scheme is insecure under KOA scenario. However it is very difficult to analyze our scheme under these scenarios.

**Table 9** The comparisons of the proposed algorithm, Liu's scheme [9], and Behina's scheme [5]

| Attribute description | | Proposed scheme | Liu's scheme[9] | Behina's scheme[5] |
|---|---|---|---|---|
| Embedding domain | | The DCT domain of the scrambled image and space domain | DWT domain | Space domain |
| Keys composition | | Embedding position key $Key1$ and encryption key and permutation key $Key2$ | Only watermark encryption key | Only embedding position key |
| Geometric attacks | Rotation (8°) | √ | × | × |
| | Translation (40 pixels) | √ | × | × |
| | Cropping (5%) | × | √ | √ |
| Cryptanalysis | WOA | Difficult | Easy access to the embedding positions | Difficult |
| | KOA | Difficult | Easy access to the embedding positions | Easy access to the embedding position |
| | KMA | Difficult | Easy access to the | Difficult |

| | embedding positions and decode the encrypted watermark |
|---|---|

*Annotations: √ indicates can resist against such an attack and × cannot.*

## 6 Conclusions

A blind watermarking scheme is presented in the paper, which enhances the security of the algorithm by protecting the watermark information and the embedding positions. For realization of the secrecy of embedding positions, we have proposed a sub key generation (SKG) mechanism that can generate different sub keys for different images. During the embedding process I, a logo watermark is embedded in the DCT domain of the scrambled image. During the embedding process II, a binary chaotic sequence called the oriented watermark is hidden in the spatial domain. Simulation results show that the proposed scheme has good invisibility and good robustness. In addition, they can resist the rotation attack and the translation attack. Moreover, a number of simulation results and analysis, such as key length analysis, key sensitivity analysis, cryptanalysis under WOA / KOA / KMA scenarios and comparison with other algorithms, demonstrate that the scheme achieve a high level of security.